\documentclass[aps,pre,twocolumn,superscriptaddress]{revtex4-1}
\usepackage{amsmath}
\usepackage{amssymb}
\usepackage[utf8]{inputenc}
\usepackage{graphicx}
\usepackage[english]{babel}
\usepackage{tikz}

\newcommand{\dd}{\; \mathrm{d}}

\newcommand{\bs}[1]{\boldsymbol{#1}}

\newcommand{\dint}[1]{\mathrm{d}{#1}}

\begin{document}

\title{Nearest-neighbor connectedness theory: A general approach to continuum percolation}

\author{Fabian Coupette}
\affiliation{Institute of Physics, University of Freiburg, Hermann-Herder-Stra{\ss}e 3, 79104 Freiburg, Germany}
\author{Ren\'e de Bruijn}
\email{r.a.j.d.bruijn@tue.nl}

\author{Petrus Bult}
\affiliation{Department of Applied Physics, Eindhoven University of Technology, P.O. Box 513, 3500 MB Eindhoven, Netherlands}
\author{Shari Finner}
\affiliation{Department of Applied Physics, Eindhoven University of Technology, P.O. Box 513, 3500 MB Eindhoven, Netherlands}
\author{Mark A. Miller}
\affiliation{Department of Chemistry, Durham
University, South Road, Durham DH1 3LE, United Kingdom}
\author{Paul van der Schoot}
\affiliation{Department of Applied Physics, Eindhoven University of Technology, P.O. Box 513, 3500 MB Eindhoven, Netherlands}
\author{Tanja Schilling}
\email{tanja.schilling@physik.uni-freiburg.de}
\affiliation{Institute of Physics, University of Freiburg, Hermann-Herder-Stra{\ss}e 3, 79104 Freiburg, Germany}

\date{\today}

\begin{abstract}
  We introduce a method to estimate continuum percolation thresholds and illustrate its usefulness by investigating geometric percolation of non-interacting line segments and disks in two spatial dimensions. These examples serve as models for electrical percolation of elongated and flat nanofillers in thin film composites. While the standard contact volume argument and extensions thereof in connectedness percolation theory yield accurate predictions for slender nanofillers in three dimensions, they fail to do so in two dimensions, making our test a stringent one. In fact, neither a systematic order-by-order correction to the standard argument nor invoking the connectedness version of the Percus-Yevick approximation yield significant improvements for either type of particle. Making use of simple geometric considerations, our new method predicts a percolation threshold of $\rho_c l^2 \approx 5.83$ for segments of length $l$, which is close to the $\rho_c l^2 \approx 5.64$ found in Monte Carlo simulations. For disks of area $a$ we find $\rho_c a \approx 1.00$, close to the Monte Carlo result of $\rho_c a \approx 1.13$. We discuss the shortcomings of the conventional approaches and explain how usage of the nearest-neighbor distribution in our new method bypasses those complications. 
\end{abstract}

\maketitle

The electrical, thermal and mechanical properties of polymeric materials can be controlled by the addition of (conductive) nanofillers, producing what may be called functional nanomaterials. For fillers to strongly affect the properties of the polymeric host material, a material-spanning network of connected particles is required. Here, connectivity is defined in terms of a length scale below which the particles are able to effectively exchange charge carriers, heat, or other quantities to be transported. The formation of such a  network is a geometrical transition akin to a phase transition,
and the set of conditions for which this occurs is referred to as the percolation threshold. Commonly used fillers are, for instance, metallic nanowires and carbon nanotubes, \cite{Guo_2013,Pasquier_2005} whose highly elongated shape is known to be ideal for producing the required material spanning network at very low filler fractions \cite{balberg_1984}. 

A wide range of possible application areas is currently under investigation, including mechanical stress sensing, actuation, energy harvesting, electromagnetic interference shielding and opto-electronics in the form of transparent thin film electrodes \cite{Choi_2019,Jiang_2019,Huang_2020,Mclellan_2020,Wang_2012,Valasma_2020,Cheng_2020}. 
In many of these application areas, it is crucial for the material to have barely crossed over from an insulating to a conductive state, either in order to keep the particle loading as low as possible or to maximize the response to an external stimulus. It stands to reason that a deeper theoretical understanding of the percolation threshold should be conducive to a rational design of novel composite materials.

Our theoretical understanding of percolation of slender nanofillers in three-dimensional bulk materials has improved significantly over the past decades since the pioneering works of Bug et al.~\cite{Bug_1985, *Bug_1986} and of Balberg et al.~\cite{balberg_1984}, in particular for particles that interact via a harshly repulsive excluded volume \cite{Schilling_2015}. This is not the case for percolation of slender particles in quasi-two-dimensional materials, that is, composite films of which the height is much smaller than the length of the filler particles \cite{Moradi_2015,Mutiso_2013}.
These quasi-two-dimensional materials are often modelled as two-dimensional systems, consisting of non-interacting fillers that are defined to be connected if they overlap. 

Even for such an ideal model system, applying the common contact volume argument to pinpoint the percolation threshold is highly inaccurate in two dimensions, despite it being asymptotically exact for infinitely slender fillers in three dimensions.
As the contact volume argument neglects direct connections between more than two neighboring particles, this approximation is similar in spirit to that of Onsager theory for the isotropic-to-nematic phase transition of hard rod-like particles, which fails to accurately describe the same phase transition in two dimensions \cite{Kayser_1978, Bates_2000}.

In this Letter we show that neither a systematic expansion in increasing powers of the density nor the connectedness version of Percus-Yevick (cPY) approximation significantly improve predictions obtained within the second virial approximation. To remedy this, we propose a novel method that is conceptually simple, yet has a significantly improved accuracy in predicting the percolation threshold, in particular for systems in which the percolating cluster contains large loops of connected particles. As the method is similar to connectedness percolation theory (CPT) but uses the nearest-neighbor distribution as input, we dub it Nearest-Neighbor Connectedness Percolation Theory (NNCPT).

NNCPT uses neither a density expansion nor a closure relation taken from liquid state theory. Instead, closure is obtained by geometrical arguments regarding the cluster structure at short distances, while the behavior at long distances is obtained by a renormalization-group-type argument. The method is exact for one-dimensional systems and reproduces the correct low density limit in higher dimensions. The computation of the percolation threshold is numerically simple, and the framework can easily be applied to arbitrary pair interactions if the corresponding nearest-neighbor distribution is inserted. Thus, our method provides a reliable way to estimate percolation thresholds for systems where conventional approaches are known to struggle -- two-dimensional ideal particles forming excellent benchmark systems. 

Given the success of CPT in predicting percolation thresholds of a large class of three-dimensional systems, we begin by recalling its basic notions and then apply them to the two-dimensional case:
The pair connectedness function $P(1,2)$ is defined such that $\rho(1)\rho(2) P(1,2) \mathrm{d} 1 \mathrm{d} 2$ is the probability that particle $1$ and particle $2$ are part of a cluster of connected particles \cite{Coniglio_1977}. Here, $\rho(1)$ and $\rho(2)$ are  single-particle number densities, the labels $1$ and $2$ represent the degrees of freedom of the particles (including orientations), and $\mathrm{d} 1$, $\mathrm{d} 2$ are the corresponding phase space volumes \cite{Hansen_2013}. For $P(1,2)$, a connectedness equivalent to the Ornstein-Zernike (cOZ) equation can be obtained:
\begin{equation}\label{eq:OZnormal}
P(1,2) = C^+(1,2) +  \int \mathrm{d} 3 \;\rho(3) C^+(1,3) P(3,2),
\end{equation}
where $C^+(1,2)$ is the so-called direct connectedness function.
The mean cluster size $S$ is given by
\begin{equation}
S = 1 + \lim\limits_{q \to 0}\rho \langle\widehat{P}(q, \vartheta_{12})\rangle = \lim\limits_{q \to 0} (1- \rho\langle\widehat{C}^+(q,\vartheta_{12}) \rangle)^{-1},
\end{equation}
where $\rho$ is the average particle number density, the hat indicates a Fourier transform, $q$ denotes the magnitude of the wave vector, $\vartheta_{12}$ is the relative orientation between the particles, and the brackets $\langle \cdots \rangle$ describe the angular average over the orientations of both particles. 
The final expression applies only for translationally and rotationally invariant distributions but is independent of particle shape \cite{Drwenski_2017}.  The percolation threshold is given by the density $\rho_\mathrm{c}$ for which $S$ diverges. 

Hence, we only require $C^+(1,2)$ to calculate the percolation threshold. This direct connectedness function is, however, typically not known in closed form. 
The simplest method to approximate it is to make use of a series expansion in the density, similar to the virial expansion describing thermodynamic properties of dispersions \cite{Coniglio_1977} 
\begin{equation}\label{eq:virial}
\lim\limits_{q \to 0} \widehat{C}^+(q) = \sum_{n = 0}^{\infty} \rho^{n} C^+_{n+2}\; .
\end{equation}
To describe percolation of three-dimensional slender particles, the expansion of Eq.~(\ref{eq:virial}) can be truncated after the $C^+_2$-term. This truncation, known as the `second connectedness virial' approximation, corresponds to the assumption that the cluster has a tree-like structure, and, as it yields mean-field critical exponents, can be considered a mean-field approximation in percolation theory. While believed to be asymptotically exact for infinitely slender particles in three dimensions \cite{Bug_1985, *Bug_1986}, its prediction for two-dimensional line segments is far less accurate. Using the analytical expression $C^+_2 = 2 l^2/\pi$ for line segments of length $l$ yields $\rho_\mathrm{c} l^2 = \pi/2 $, which deviates by a factor of $3.6$ from Monte Carlo simulation results \cite{Mertens_2012}. The situation is even worse for disks, where the disparity is a factor of 4.5 (see Table \ref{tab:virial}). 
To improve upon this unsatisfactory situation one could, perhaps naively, either extend the estimate of $C^+$ in an order-by-order fashion, or make use of the extensive toolbox of closures obtained from liquid-state theory \cite{Coniglio_1977,Torquato_2002}. 
 
\begin{figure}[tb]
	\centering
	\includegraphics[width=0.99\columnwidth]{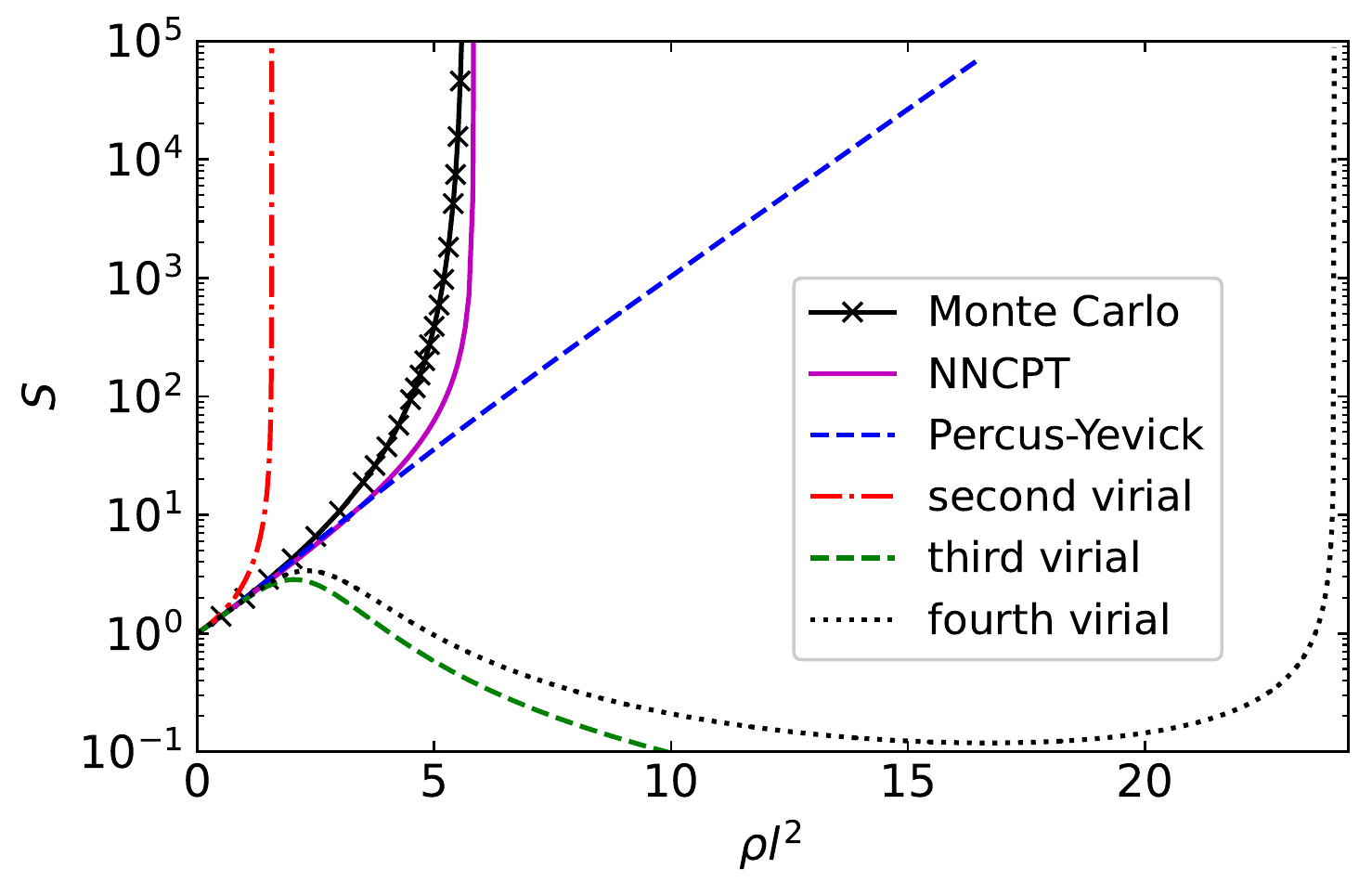}
	\caption{Mean cluster size of overlapping line segments in two dimensions. The third and fourth virial approximations give non-monotonic curves, and the cPY prediction does not diverge for densities below $\rho l^2 = 17$.} 
	\label{fig:preav}
\end{figure}

For the order-by-order approach, we use the known analytical expression for the third hard-body virial coefficient $B_3$ for line segments \cite{Tarjus_1991}, which can be linked to $C_3^+$ by $C_3^+ = - 3 B_3$. 
Truncating after $C^+_3$ produces two imaginary and hence unphysical predictions for the percolation threshold. Higher order terms can be obtained by Monte Carlo integration. Truncating after $C^+_4 l^{-6} = (0.00548 \pm 0.00002)$ we obtain two imaginary roots and one real root, $\rho_\mathrm{c} l^2 = 24.0 \pm 0.1 $. Hence, going up to fourth order does not provide any improvement upon the second virial approximation. Moreover, the mean cluster size $S$ obtained within the third and the fourth virial approximation produces a non-monotonic function of the scaled density, see Fig.~\ref{fig:preav}. $S$ even becomes smaller than unity, indicating negative connectivity probabilities and highlighting the problems with the virial expansion. The same happens for overlapping disks
(see Table~\ref{tab:virial}).

\begin{table}[tb]
	\centering
	\caption{Percolation threshold $\rho_c $ obtained from Monte Carlo simulations \cite{Mertens_2012}, the new method NNCPT, the virial expansion up to fourth order, the connectedness Percus-Yevick theory \cite{deBruijn_2020} and a Padé approximant \cite{Torquato_2012} in the mean cluster size for overlapping lines of length $l$ and disks of area $a$.}
	\label{tab:virial}
	\begin{tabular*}{3.4in}{c c c}
		\hline\hline
		Method & line segments ($\rho_\mathrm{c} l^2$) &  disks ($\rho_\mathrm{c} a$)\\
		\hline
		Monte Carlo & $5.6372858(6)$ & $1.12808737(6)$ \\
		NNCPT & 5.83 & 1.00 \\
		Second Virial & $\pi/2$ $\approx 1.57$ & $\frac{1}{4}$ \\
		Third Virial & - & -  \\
		Fourth Virial & 24.0 ($\sigma  = 0.1$) & $0.326838\dots$ \\
		cPY & - & - \\
		$[2,1]$-Padé approximant & $3.878$ ($\sigma = 10^{-3}$) & $0.748742\dots$ \\
		\hline\hline
	\end{tabular*}
\end{table}

Rather than adding virials, we next invoke the connectedness Percus-Yevick (cPY) closure \footnote{For ideal particles, this closure is defined by the exact identity $P(1,2) = 1$ for overlapping particles and the approximate identity $C^+(1,2) = 0$ if the particles do not overlap.}. If expressed in terms of a diagrammatic expansion of the pair connectedness function, all diagrams representing the second and third virials are contained exactly, but higher order ones are incomplete \cite{Stell_1984,DeSimone_1986}. For the thermodynamic properties of hard particles, these diagrams turn out to be unimportant. However, as we shall see, they are important in the context of percolation in systems in which the percolating cluster contains large loops.  

The cPY closure needs to be solved self-consistently in conjunction with the cOZ equation. 
We tackle this numerically using a rotational invariant expansion, see appendix.  We iterate the governing set of equations by a modified Picard iteration \cite{Talman_1978,Hamilton_2000} for all densities up to the one where the iterations no longer converge and percolation is achieved. As shown in Fig.~\ref{fig:preav}, there is neither a percolation threshold for line segments nor for disks \cite{deBruijn_2020}.

\begin{figure}
	\includegraphics[width=0.49 \textwidth]{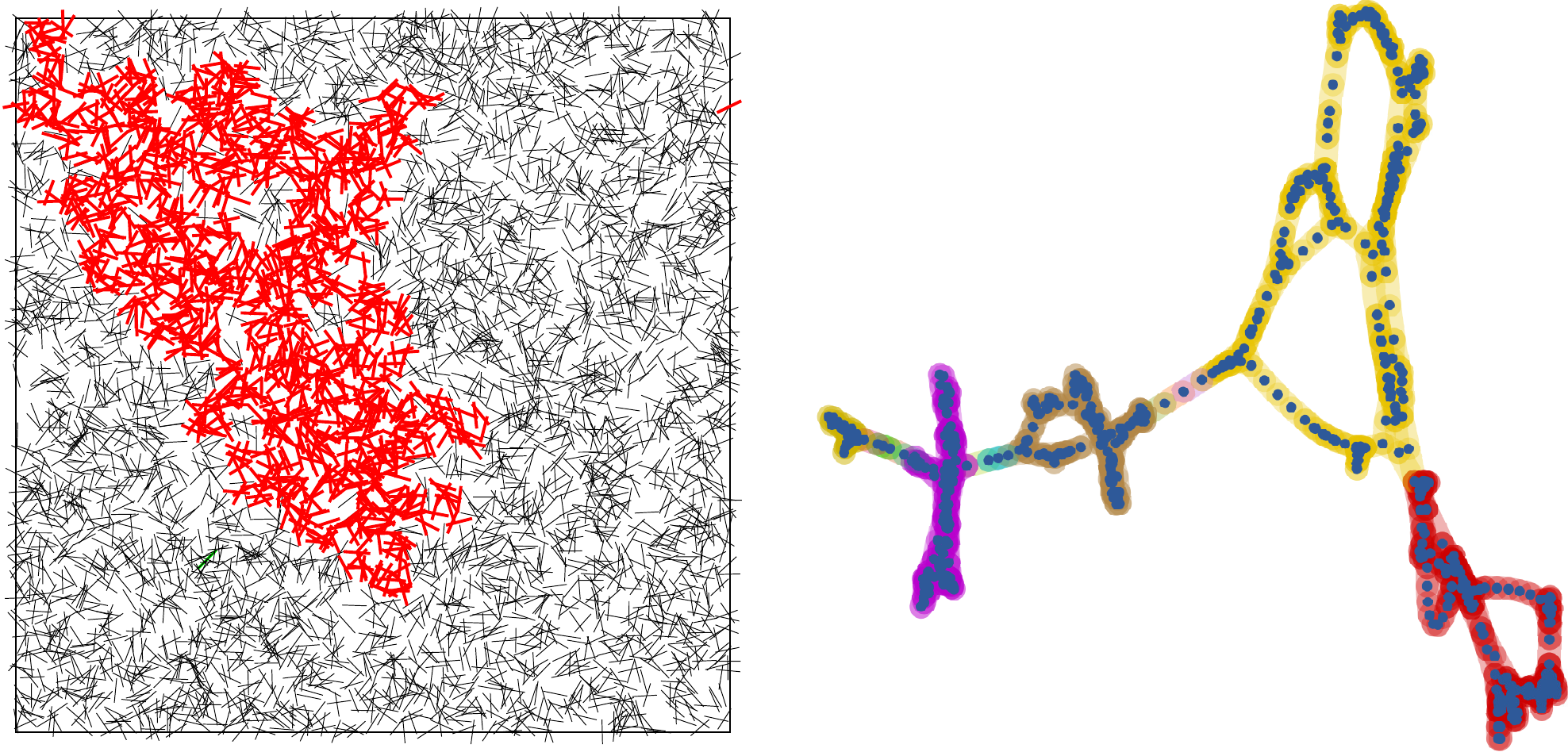}
        \caption{Left: Simulation snapshot at $\rho l^2 = 5.0$; the largest connected cluster is marked in red. Right: Backbone of the network connecting two rods within the largest cluster. Colors correspond to different non-nodal components. Some of these contain large loops, hence a virial expansion of the direct connectedness function is expected to fail.}
        \label{2dSnapshots}
\end{figure}

Hence we suggest that methods borrowed directly from liquid state theory must be inherently inaccurate when applied to percolation problems. We attribute the failure of these methods to ignoring loop correlations involving very large numbers of particles. To test for the occurrence of such loops, we carried out a simulation, in which we dropped line segments on the plane randomly with positions and orientations drawn uniformly. We then identified all clusters of intersecting line segments and counted the loops in the backbone of each cluster. In Fig.~\ref{2dSnapshots} we show a snapshot at a density just below the percolation transition. On the right we show the structure of the backbone of the largest cluster. The various non-nodal components ({\em i.e.,} sections where no single particle lies on all paths between the end-points) are colored differently. Large loops consist of many particles, and can therefore not be captured by a low order expansion, as the $n^\mathrm{th}$ coefficient $C_n^+$ in the truncated virial expansion describes loops consisting of at most $n$ particles. This suggests that a different theoretical approach is needed altogether.

  \begin{figure}[tb]
  \centering
    \includegraphics[width=0.8\columnwidth]{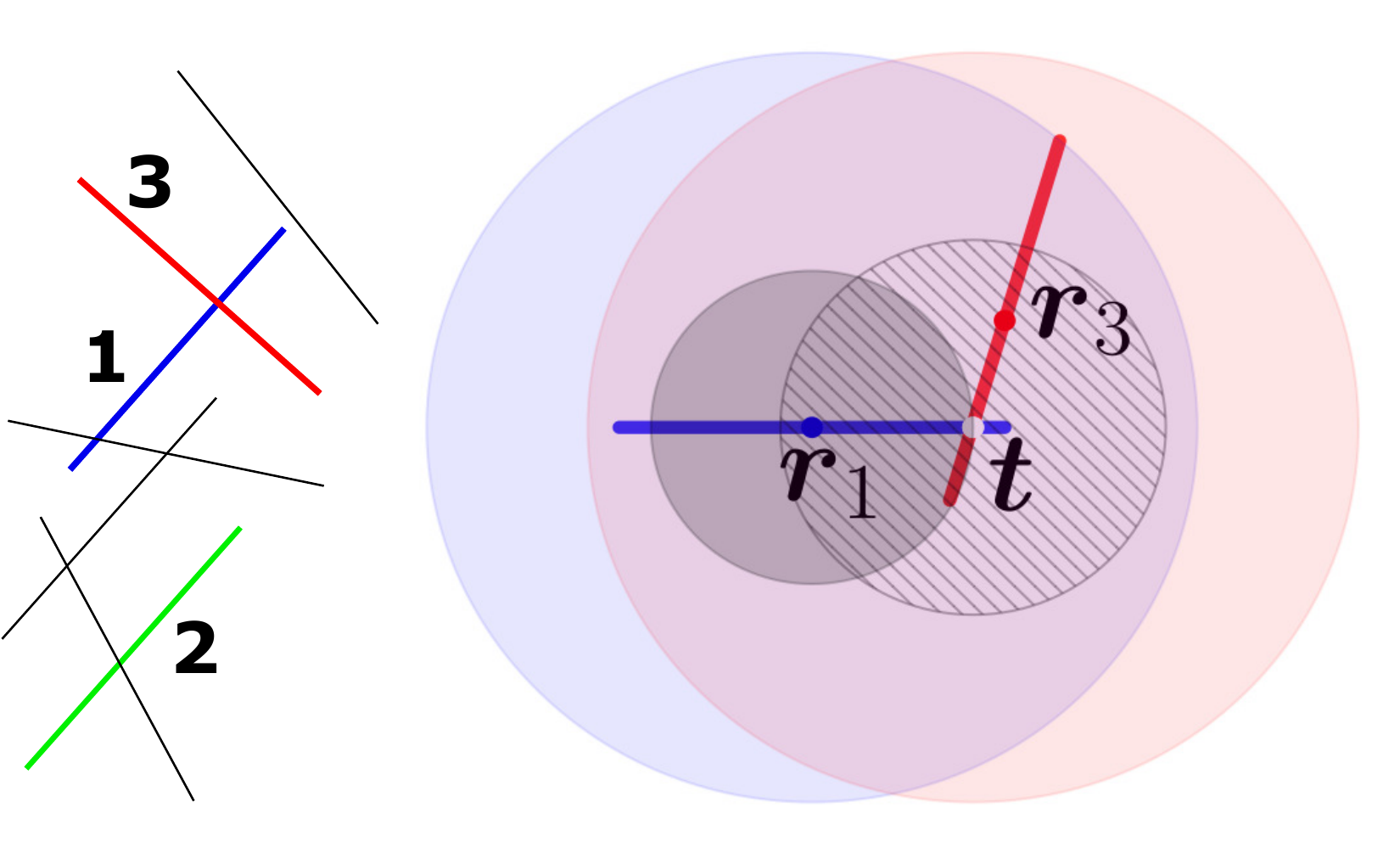}
    \caption{Left: Construction of the pair connectedness function in NNCPT. Right: Effective treatment of the union of a segment and its nearest neighbor as weighted disks: The hatched disk represents the nearest neighbor, the large transparent disks form the \textit{surface} $A_{1 \cup 3}$ with exception of the dark disk which is blocked by the nearest neighbor condition.}
    \label{fig:Skizze}
  \end{figure}
  
In contrast to the work by Coniglio \cite{Coniglio_1977}, we construct the pair connectedness function $P(1,2)$ iteratively by going going from particle $1$ to its nearest neighbor (particle 3 in Fig.~\ref{fig:Skizze}). If particle 1 is not connected to its nearest neighbor, it cannot be part of any connected cluster. If it is connected, we treat the union of both particles as a ``new particle 1'' and repeat the process until we reach particle $2$. Thus $P(1,2)$ can be written as
\begin{eqnarray}
  P(1,2) &=& f^+(1,2)  \label{eq:keytar} \\
  &+&\nonumber \left(1- f^+(1,2) \right) \int \dd 3 \, \omega^+(1,3)  P(3,2|3n1). 
\end{eqnarray}
Here, $f^+$ denotes the probability that particles 1 and 2 are directly connected. $\omega^+$ is the probability density to encounter the nearest neighbor of 1 in the phase space volume $\rm{d}3$, while also being connected to it. ($\omega^+$ differs from $f^+$, as $f^+$ refers to any connection, while $\omega^+$ only refers to the connection with the nearest neighbor. Both functions need to be defined suitably for a given particle geometry.) $P(3,2|3n1)$ is the conditional probability that 3 and 2 are part of the same cluster, given that 3 is the nearest neighbor of 1. 

Compared to Eq.~(\ref{eq:OZnormal}), Eq.~(\ref{eq:keytar}) bears the disadvantage that it connects two different types of probability. We hence define the ratio $c(3,2|3n1):=\frac{P(3,2|3n1)}{P(3,2)}$ and obtain our alternative to the cOZ equation, the nearest neighbor connectedness equation
\begin{eqnarray}
  &P(1,2) = f^+(1,2)&  	\label{eq:hotkey} \\
  &+ 
  \left(1- f^+(1,2) \right) \int \dd 3 \, \omega^+(1,3) c(3,2|3n1) P(3,2).& \nonumber 
\end{eqnarray}
Here, $f^+$ is a known function, which depends on the specific geometry of the problem, while $P$, $\omega^+$ and $c$ are in general not known. However in contrast to the direct connectedness function in Eq.~(\ref{eq:OZnormal}), which involves an infinite sum of arbitrarily complicated diagrams, the nearest neighbor distribution $\omega^+$ can be approximated by geometric considerations, and a closure for Eq.~(\ref{eq:hotkey}) can be obtained by specifying the conditional probability $c$. Moreover, if we assume $c$ to be independent of particle 2, Eq.~(\ref{eq:hotkey}) is reduced to a convolution. Close to percolation this is a reasonable assumption, because irrespective of the distance between particle 2 and particle 3, what matters for $P(3,2)$ is whether they are part of the percolating cluster. For one-dimensional systems this statement is actually exact \cite{Coupette_2020}. Once the problem is simplified in this way, the remaining task is to find the density at which the integral over the complete integral kernel exceeds one.

We now apply NNCPT to freely penetrable line-segments. The relative position of two segments with respect to each other is given by their orientations and their center-to-center distance $\bs{r}_{12}$. The connectivity criterion,  which determines whether two segments intersect each other, is invariant under rotations of the whole system. Thus, in place of two orientational degrees of freedom, we only need to account for the relative orientation $\vartheta_{12}$. Using translational invariance of the connectivity criterion, we are free to choose segment 1 as centered in the origin parallel to the $x$-axis. We now need to specify $f^+$, $\omega^+$ and $c$. $f^+(\bs{r}_{12},\vartheta_{12})$ is $1$ if the segments intersect each other and zero if they do not. To specify $\omega^+$, we define as the nearest neighbor the segment which intersects segment 1 closest to its center. $\omega^+$ can be expressed as a function of the intersection point $\bs{t}$ along segment 1. Note that if a segment intersects at a point $\bs{t}$, the center position of the nearest neighbor completely defines its orientation. 
The center of a segment 3, $\bs{r_3}$, which intersects segment 1 at position $\bs{t}$, can be anywhere within a circle of diameter $l$ around $\bs{t}$ with uniform probability. As a consequence, the nearest neighbor distribution becomes an exponential evenly distributed on a disk: 

\begin{align}
	\omega^+(\bs{r_3},t) &= \frac{4}{\pi l^2}\Theta \left(\frac{l}{2}-|\bs{r_3}-\bs{t}|\right)\frac{2}{\pi } \rho l \exp\left(- \frac{4 \rho l }{\pi} |t|\right) ,
\end{align}

with the number density of segments $\rho$, $t$ the projection of $\bs{t}$ on the $x$-axis, and $\Theta$ the Heaviside function. Notice that we fixed the orientation of segment 1 so that $\omega^+$ depends only on $t$ and not on $\bs{t}$.

The last function to specify is $c$, which measures how much the probability of a connection between segments 3 and 2 is influenced by the existence of segment 1. This probability is the result of two competing effects. On the one hand, two segments together offer more phase space than a single one to other segments to connect to. On the other hand, a portion of this phase space is blocked by the condition that 3 is the nearest neighbor of 1. We argue that close to the percolation threshold, the only relevant difference between a single segment and two intersecting segments is their \textit{surface} in phase space. If we choose $c$ as a function of the surface in phase space, it apparently does not depend on segment 2 anymore -- our first approximation -- but still on the detailed microscopic arrangement of segments 1 and 3. Eq.~(\ref{eq:hotkey}) thus reads,  
\begin{align}
    & P(\bs{r}_{12},\vartheta_{12}) = f^+(\bs{r}_{12},\vartheta_{12}) + (1- f^+(\bs{r}_{12},\vartheta_{12})) \; \times \label{eq:hotkeytar}  \\  \nonumber
  &\int_{-\frac{l}{2}}^\frac{l}{2} \dd t \int \dd \bs{r}_{3} \; \omega^{+}(\bs{r}_{3},t) c( \bs{r}_{3},t) P(\bs{r}_{32},\vartheta_{32}) \; .
\end{align}
Finally, we eliminate the angular dependencies. We interpret each line segment as a weighted disk, \textit{i.e.}, smeared out into all possible orientations. The contact function $f^+(1,2)$ becomes the overlap of two disks with center separation $r_{12}$ reproducing the correct angular average $f^+(r_{12}):= \langle f^+(1,2) \rangle_{\vartheta(r_{12})}$. 
In order to construct $c$, we consider the combination of two disks as a single disk with a new diameter that offers the same phase space surface, $A_3=\pi l^2$, as the union of both disks (see Fig.~\ref{fig:Skizze}). The \textit{surface} $A_{1 \cup 3}(t)$ of two disks at center-to-center separation $t$ is given by 
$
  A_{1 \cup 3}(t) =  \pi (l^2-t^2) + 2 l^2 \arcsin\left(\frac{t}{2 l} \right) +\frac{t}{2} \sqrt{4 l^2 - t^2} 
$.
Note that a disk of radius $t$ is blocked on account of the nearest neighbor constraint resulting in $-\pi t^2$.  Recall that the potential center positions of a nearest neighbor intersecting at $\bs{t}$ are homogeneously distributed on a disk of diameter $l$. For each $t$ we choose this disk as the weighted disk representing the nearest neighbor. The effective diameter thus depends only on $|t|$:
\begin{equation}
   c(|t|) :=  \frac{l_{\mathrm{eff}}(|t|)}{l} = \left(\frac{A_{1 \cup 3}(|t|)}{A_3}\right)^{\frac{1}{2}} \; .
   \label{eq:close}
\end{equation}
Referring to Eq.~(\ref{eq:keytar}), we account for the existence of segment 1 by resizing segment 3 to an effective length $l_{\mathrm{eff}}$ and make segment 3 the new segment 1. The newly formed segment is subjected to the same nearest neighbor distribution as before; only the boundaries of the $t$-integral change. The substitution $t' = t l$ reveals that $c$ as defined in Eq.~(\ref{eq:close}) has the desired effect:
\begin{align}
    & P(r_{12}) = f^+(r_{12}) + (1- f^+(r_{12})) \; \times \label{eq:numericaleq} \\  \nonumber
  &\int_{-\frac{1}{2}}^\frac{1}{2} \dd t' \; l \int \dd \bs{r}_{3} \; \omega^+(\bs{r}_{3},l t')  c(l |t'|)  P(r_{32}) \; .
\end{align}
Analogously to CPT, the percolation threshold is given by the smallest density that satisfies
\begin{align}
0 = 1 - \int \dd \bs{r}_3 \; \int_{-\frac{1}{2}}^\frac{1}{2} \dd t' \;   \, \omega^+(\bs{r}_3,l t') l_{\mathrm{eff}}(l |t'|) \; , 
\label{eq:final}
\end{align}
yielding $\rho_c l^2 \approx 5.83$, which is very close to the result of Monte Carlo simulations (cf. Table \ref{tab:virial}). A good approximation of the mean cluster size can be computed as well (see Fig.~\ref{fig:preav}). Detailed information is given in the SI. 
For intermediate densities, the NNCPT prediction slightly deviates from the simulation results. Our choice of $c(|t|$) is density-independent and designed to yield a good approximation for the percolation threshold. However, the correct kernel $c(3,2|3n1)$ depends on the density. A good approximation of $c(3,2|3n1)$ at the percolation threshold is therefore expected to be less good for different densities. The divergence of the mean cluster size is characterized by the mean field critical exponent $\gamma = 1$ if $c$ does not depend on the density. This offers a different angle for interpreting critical exponents and outlines a way to improve on our closure.
Usage of the nearest-neighbor distribution, however, provides the correct low density limit.     

We can use the same closure to treat the percolation of ideal disks of area $a$ and find $\rho_c a \approx 1.00$ which is also close to the simulation results \cite{Mertens_2012}. 
The good agreement between NNCPT and simulation hinges on the fact that the nearest-neighbor construction describes loop structures in the cluster backbone implicitly, not explicitly as CPT. Moreover, NNCPT avoids an explicit density expansion by comprising all information on thermal correlations between particles in the nearest neighbor distribution. This is particularly important if the critical density is as large as for ideal line segments. Furthermore, NNCPT bears the advantage that the kernel $c(3,2|3n1)$ can be directly observed in simulations delineating a straightforward approach to assess the accuracy of the approximations made. 

To summarize, we have introduced a novel method to calculate continuum percolation thresholds. As two-dimensional systems are particularly challenging for standard methods of percolation theory, we have studied geometrical percolation of ideal line segments and disks in two dimensions and have shown that that our method provides accurate predictions. The method can easily be applied to interacting and polydisperse systems in three dimensions as well. We therefore expect it to be of use in the design of composite materials.

\begin{acknowledgments}
F.C.~and T.S. acknowledge funding by the German Research Foundation in project 404913146. R.d.B. and P.v.d.S. acknowledge funding by the Institute for Complex Molecular Systems at Eindhoven University of Technology.
\end{acknowledgments}

\appendix

\section*{Rotationally invariant expansion}
The connectedness Ornstein-Zernike (cOZ) equation is given by
\begin{equation}\label{eq:cOZ}
P(1,2) = C^+(1,2) + \int \dd 3 \rho(3) C^+(1,3) P(3,2),
\end{equation}
where $P(1,2)$ is the pair connectedness function, $C^+(1,2)$ the direct connectedness function, and $\rho(3)$ is the single-particle number density, which in the currently relevant isotropic case is $\rho(3) = \rho/2 \pi$ in two spatial dimensions. The numbers $1$, $2$ and $3$ are particle labels, and represent the orientational and positional degrees of freedom of the particles. As stated in the main text, this equation links the two unknown functions $C^+(1,2)$ and $P(1,2)$, and must be supplemented by a closure relation for $C^+(1,2)$. Here, we use the connectedness Percus-Yevick (cPY) closure, defined by
\begin{align}
P(1,2) &= 1 &&\text{if particles $1$ and $2$ overlap, and} \\
C^+(1,2) &= 0 &&\text{if particles $1$ and $2$ do not overlap.}
\end{align}
Since we solve the set of cOZ - cPY equations self-consistently, we cannot \textit{a priori} average out the orientational degrees of freedom. To properly include the orientational degrees of freedom, we use a rotationally invariant expansion. 

We follow the approach by Ferreira and coworkers \cite{Ferreira_1991} for two-dimensional isotropic dispersions of anisometric particles, and extend it to connectedness percolation theory. We start by expanding the functions $P(1,2)$ and $C^+(1,2)$ as
\begin{equation}
P(1,2) = \sum_{m,n = -\infty}^{\infty} P^{mn}(\mathbf{r}) \Psi^{mn}(\vartheta_1,\vartheta_2),
\end{equation}
where $P^{mn}(\mathbf{r})$ are the projections of $P(1,2)$ on the rotationally invariant basis defined by $\Psi^{mn}(\vartheta_1,\vartheta_2)$, which, by definition, remain invariant under rotations of the whole system. Using the inter-molecular reference frame, which corresponds to describing the orientations of the particles with respect to their separation vector $\mathbf{r}$, we obtain 
\begin{equation}
\Psi^{mn}(\vartheta_1,\vartheta_2) = \exp i (m \vartheta_1 + n \vartheta_2).
\end{equation}
Here, $\vartheta_1$ and $\vartheta_2$ are the angles between the separation vector $\mathbf{r}$ and the major axis of the particles $1$ and $2$, respectively. These basis functions are orthogonal, and therefore the projections $P^{mn}(\mathbf{r})$ are given by 
\begin{equation}
 P^{mn}(r) = \frac{1}{4 \pi^2}\int_0^{2\pi} \dd \vartheta_1 \int_0^{2\pi}\dd \vartheta_2 P(1,2)\Psi^{mn}(\vartheta_1,\vartheta_2).
\end{equation} 
We can reduce the number of independent coefficients by noting that $P$ and $C^+$ are real functions, and using the symmetry of the particle interactions, which gives $P^{mn} = P^{-m-n} = P^{nm} = P^{-n-m}$. Secondly, the inversion symmetry of the particles demands that $m,n$ are even.

Since we solve the cOZ Equation~(\ref{eq:cOZ}) in Fourier space, we introduce the Fourier transforms of $P(1,2)$ and $C^+(1,2)$ as
\begin{equation}
\widehat{P}(\mathbf{q}, \vartheta_{1q}, \vartheta_{2q}) = \int \dint{\mathbf{r}} e^{i \mathbf{q}\cdot \mathbf{r}} P(1,2),
\end{equation}
where $\vartheta_{1q, 2q}$, are the angles between the symmetry axis of particles $1$ or $2$ and the direction of the wave vector $\mathbf{q}$.  Hence, we can relate the Fourier-space projections and the real-space projections through the Fourier-Bessel transform
\begin{equation}
\widehat{P}^{mn}(q) = 2 \pi i^{m+n} \int \dint{r} \, r P^{mn}(r) J_{m+n}(qr),
\end{equation}
where $J_N$ is the Bessel function of order $N$ \cite{Ferreira_1991} and $q = |\mathbf{q}|$ the magnitude of the wave vector. Remaining in Fourier space, we use the convolution theorem to obtain a projection-coupled cOZ equation
\begin{equation}
\widehat{P}^{mn}(q) = \widehat{C}^{mn}(q) + \rho \sum_l \widehat{P}^{ml}(q)\widehat{C}^{-ln}(q),
\end{equation}
where we have written $C$ instead of $C^+$ for clarity. Here, the orientational degrees of freedom are only implicitly present in the projections. 

We can follow the same approach for the (real-space) cPY closure, which can be written as 
\begin{equation}
C^+(1,2) =  f^+(1,2)\left[1 - P(1,2) + C^+(1,2)\right],
\end{equation}
where $f^+(1,2)$ is the contact function that is defined as $f^+(1,2) = 1$ if particles $1$ and $2$ overlap, \textit{i.e.}, are connected, and $f^+(1,2) = 0$ if particles $1$ and $2$ do not overlap. Using the same expansion, we obtain
\begin{equation}
  \begin{split}
    C^{mn}(r) = \sum_{m',n'= - \infty}^{\infty}\left[\delta_{n'0}\delta_{m'0}-\left(P^{m'n'}(r)-C^{m'n'}(r)\right)\right]\\ \times {f^+}_{n-n'}^{m-m'}(r),
    \end{split}
\end{equation}
where ${f^+}_{n-n'}^{m-m'}(r)$ are the projections of $f^+(1,2)$, but using a sub- and superscript instead of double superscript notation for clarity. The projections of $f^+(1,2)$ can be calculated using the overlap criterion yielding
\begin{equation}
{f^+}_{n-n'}^{m-m'}(r) = \frac{1}{4 \pi^2} \int_{\mathrm{Overlap}} \dint\vartheta_1 \dint{\vartheta_2} \Psi^{mn}(\vartheta_1,\vartheta_2)\Psi^{m'n'}(\vartheta_1,\vartheta_2),
\end{equation}
and must generally be calculated numerically. These projections need only be calculated once for a given model, which simplifies the self-consistent calculation of the cOZ equation. The double integral ${f^+}^N_M(r)$ was simplified by Ferreira and co-workers \cite{Ferreira_1991} to 
\begin{equation}
  \begin{split}
    {f^+}^M_N(r) = \frac{4}{\pi^2} \int_0^{\pi/2} \dint{\vartheta_1} \cos\left[M \vartheta_1 + \frac{N}{2}(\vartheta_{2\mathrm{a}} + \vartheta_{2\mathrm{b}}) \right]\\ \times \frac{\sin\left[\frac{N}{2}\left(\vartheta_{2\mathrm{a}} - \vartheta_{2\mathrm{b}}\right)\right]}{N},
    \end{split}
\end{equation}
where $\vartheta_{2a}$ and $\vartheta_{2b}$ are defined as the bounds for the angles of particle $2$ between which the particles overlap for fixed separation distance $r$ and angle $\vartheta_1$. For line segments, these angles are given by
\begin{align}
\vartheta_{2\mathrm{a}} &= \begin{cases}
\arccos{\frac{\cos{\vartheta_1}-2x}{\sqrt{1+4x^2-4 x \cos{\vartheta_1}}}}  \qquad &\text{for $\vartheta_1 \leq \arccos{x}$,}\\
\vartheta_1 +\arcsin{2 x \sin{\vartheta_1}} \qquad &\text{for $\vartheta_1 > \arccos{x}$,}
\end{cases} \\
\vartheta_{2\mathrm{b}} &=  \pi + \vartheta_1 - \arcsin\left[2 x \sin\vartheta_1\right],
\end{align}
with $x = r/l$ \footnote{For line segments, these angles can be derived as follows: Note that the edges of the overlap region always correspond to the \textsl{tip} of one of the line segments touching the other line segment somewhere. Using this, the parametric representation of line segments and the inversion symmetry of particles, we find several solutions that are easily checked for consistency with the overlap criterion.}.

Finally, the mean cluster size as defined in the main text can be written as
\begin{equation}
S = 1 + \rho \lim\limits_{q \to 0} P^{00}(q) = \lim\limits_{q \to 0}\left(1- \rho  \widehat{C}^{00}(q)\right)^{-1}
\end{equation}
in terms of the $m, n = 0$ projection only, which indeed corresponds to the isotropically averaged $P(1,2)$ and $C^+(1,2)$. Here, we have used the fact that for $q \to 0$, the projection-coupled cOZ equation can be shown to decouple \cite{Ferreira_1991}.

\section*{Ideal line segments}
\begin{figure}[tb]
	\centering
	\includegraphics[width=0.9\columnwidth]{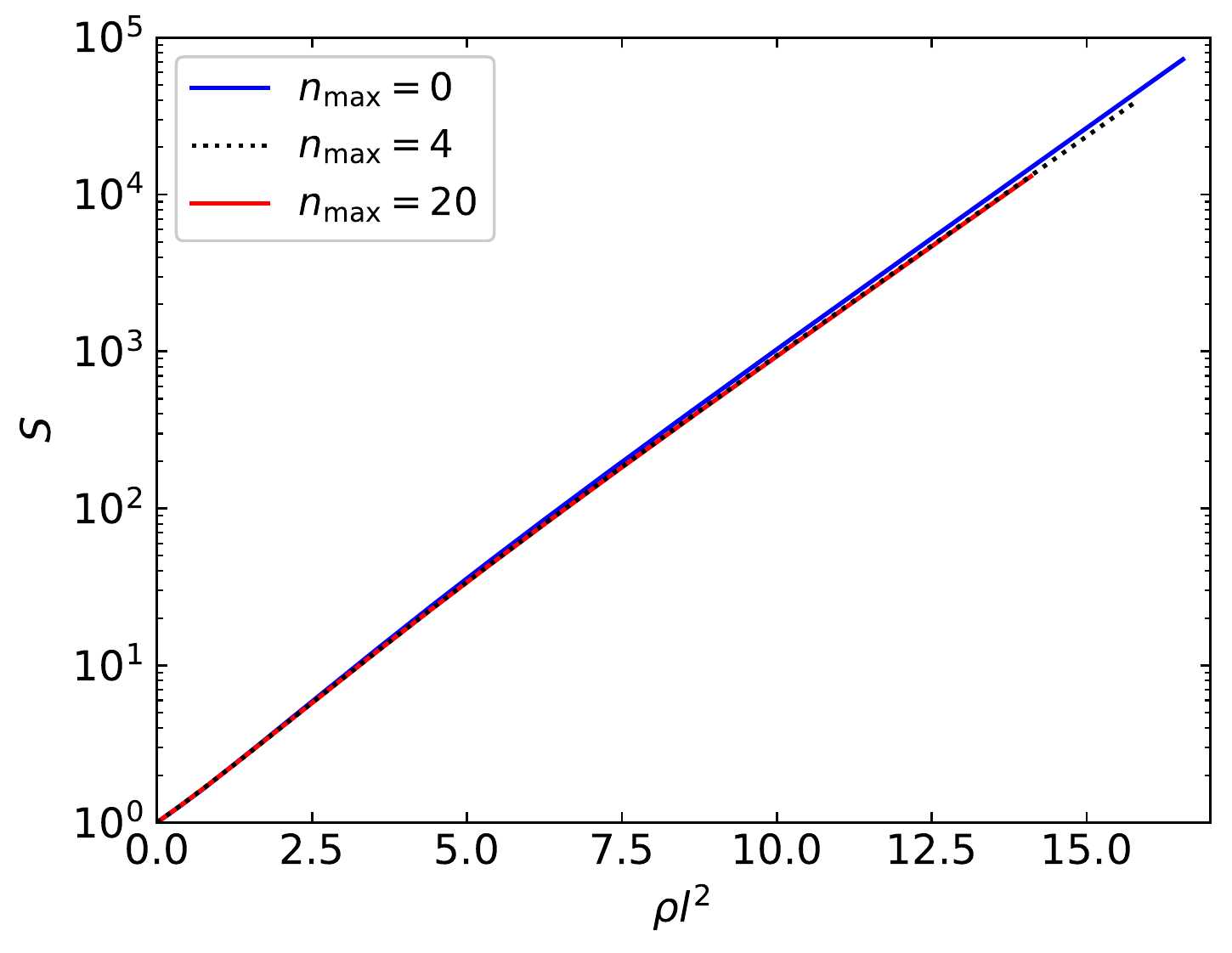}
	\caption{The mean cluster size $S$ of ideal line segments of length $l$ as function of the reduced density $\rho l^2$, obtained within the cPY closure. We include the values where the rotational invariant expansion is truncated at a different max values $n_\mathrm{max} = 0$ (blue), $n_\mathrm{max} = 4$ (black, dotted) and $n_\mathrm{max} = 20$ (red).}
	\label{fig:mcsrotinv}
\end{figure}

We solve the set of equations given by the cOZ and cPY equation self-consistently using an iterative solver. This iterative solver treats the cOZ equation in Fourier space, while dealing with the cPY closure in real space. The Fourier-Bessel transforms are efficiently handled using a method based on a logarithmic grid spacing \cite{Talman_1978,Hamilton_2000}. While the rotationally invariant expansion is derived for an infinite number of basis functions $\Psi^{mn}(\vartheta_1,\vartheta_2)$, we limit it to a finite and preferably small number. Specifically, we truncate the expansion set at $|\mathrm{max}(m,n)| = n_\mathrm{max}$. Note that $n_\mathrm{max} = 0$ corresponds to a pre-averaged approximation, \textit{i.e.}, replacing $P(1,2)$, $C^+(1,2)$, and $f^+(1,2)$ by weighted disks. For detailed orientational information, we use a relatively large value of $n_\mathrm{max} = 20$, whereas for orientationally averaged quantities, such as the mean cluster size $S$, we can truncate it at a small value. To confirm that the mean cluster size $S$ is indeed only weakly dependent on $n_\mathrm{max}$, we plot it for various values for $n_\mathrm{max}$ in Fig.~\ref{fig:mcsrotinv}. While the deviation in the $n_\mathrm{max} = 0$ approach compared to a $n_\mathrm{max} = 20$ approach is visible, it relatively minor, approximately $10\%$ at $\rho l^2 = 14.1$. For $n_\mathrm{max} = 4$, the difference at the same density is only slightly more than one per cent. Due to the minor influence of $n_\mathrm{max}$ on the mean cluster size $S$, we argue that even the pre-averaged cPY approach ($n_\mathrm{max} = 0$) yields quantitative predictions within the cPY approximation. Fig.~\ref{fig:mcsrotinv} shows that the cPY approximation does not yield a reasonable percolation threshold as we discuss in the main text, and that this observation is not altered by the choice of $n_{\mathrm{max}}$. 

\begin{figure}[tb]
	\centering
	\includegraphics[width=0.48\columnwidth]{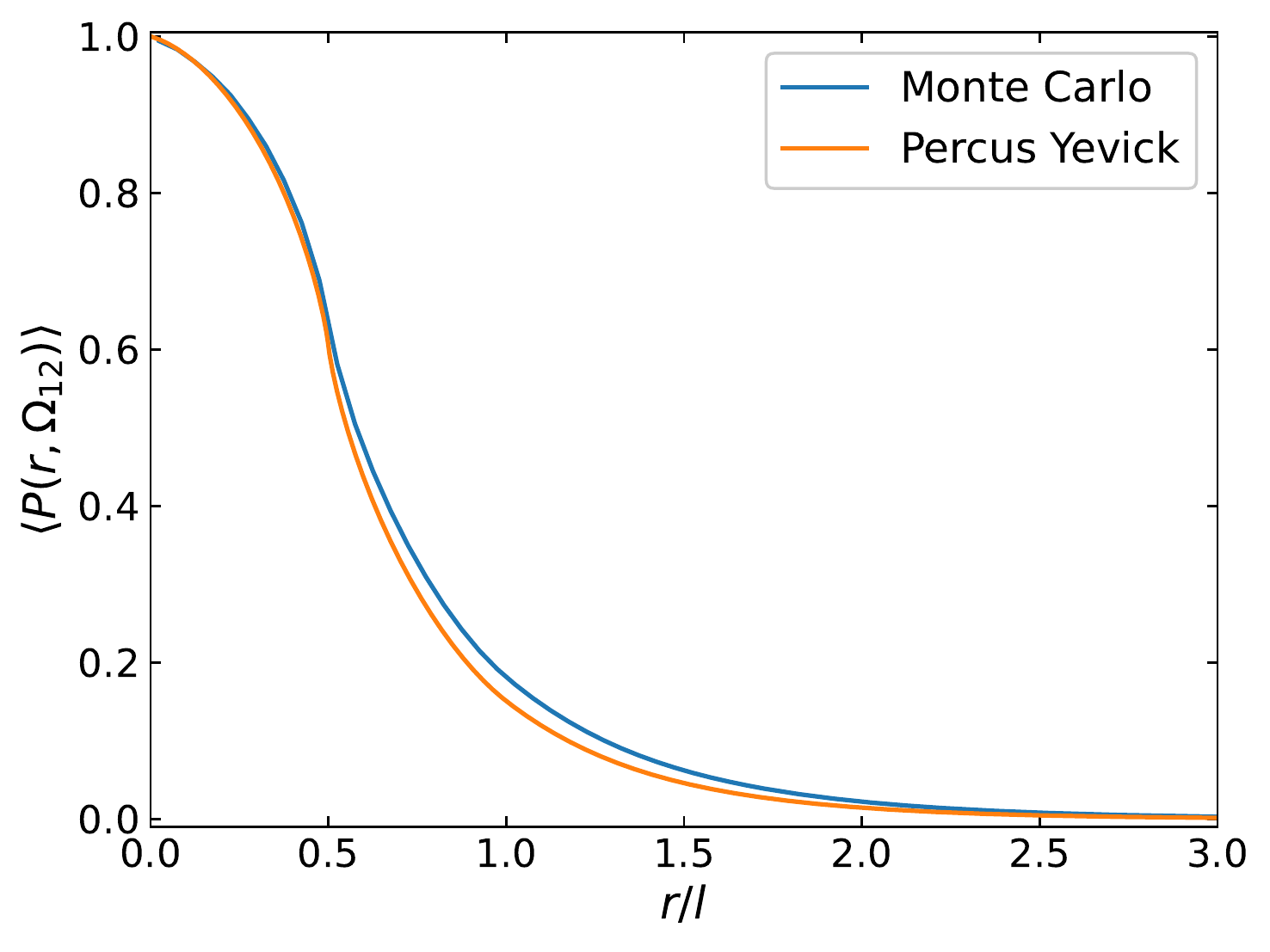}
	~
	\includegraphics[width=0.48\columnwidth]{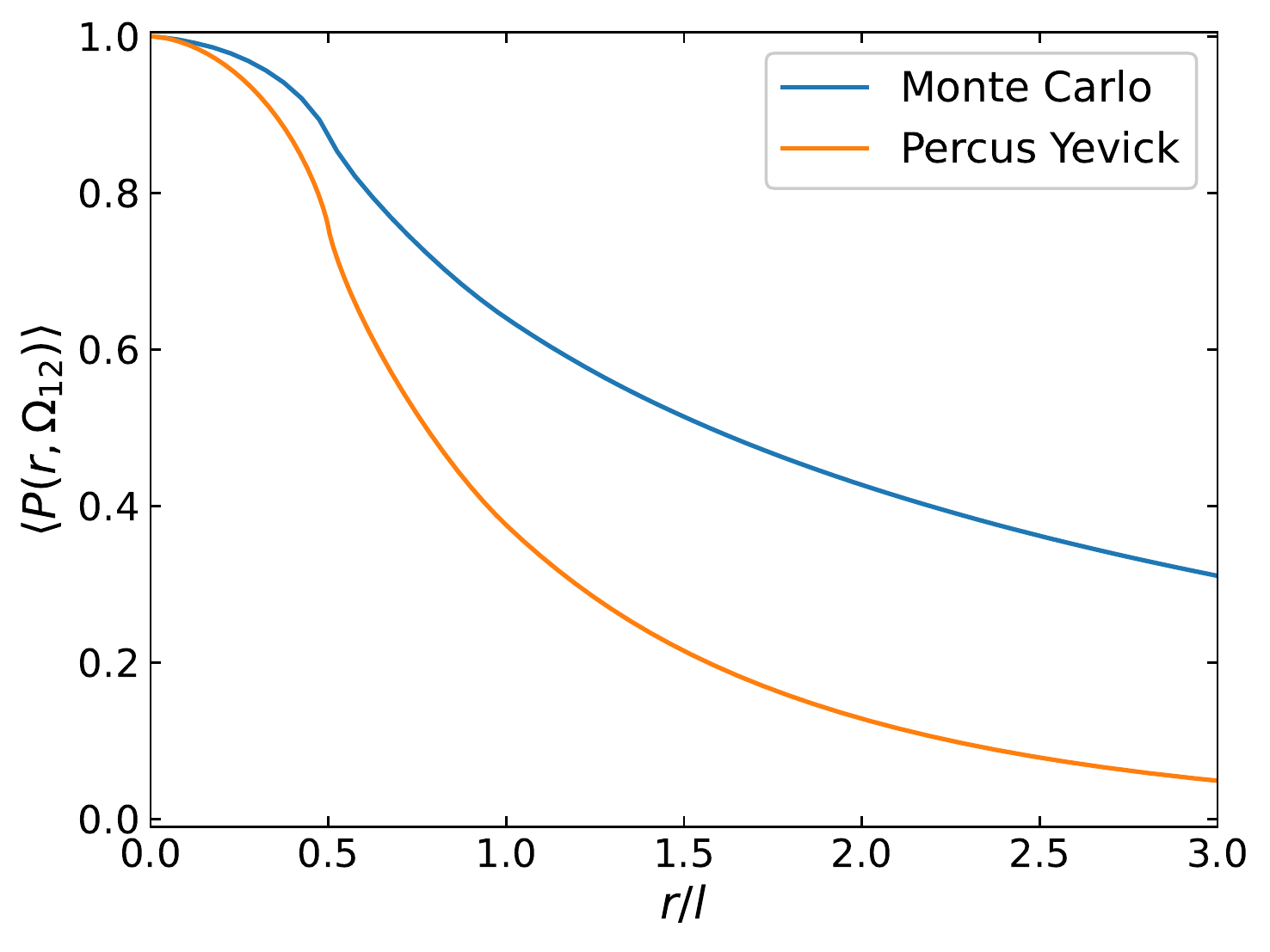}
	\caption{The orientationally averaged pair connectedness function $\langle P(r,\vartheta)\rangle$ as function of the scaled center-to-center distance $r/l$ of line segments of length $l$, for scaled densities $\rho l^2 = 2.5$ (left) and $\rho l^2 = 5.0$ (right). Here, we show both our theoretical prediction based on the cPY approximation, and the results from our Monte Carlo simulations.}
	\label{fig:pcf}
\end{figure}

\begin{figure}[tb]
	\centering
	\includegraphics[width=0.48\columnwidth]{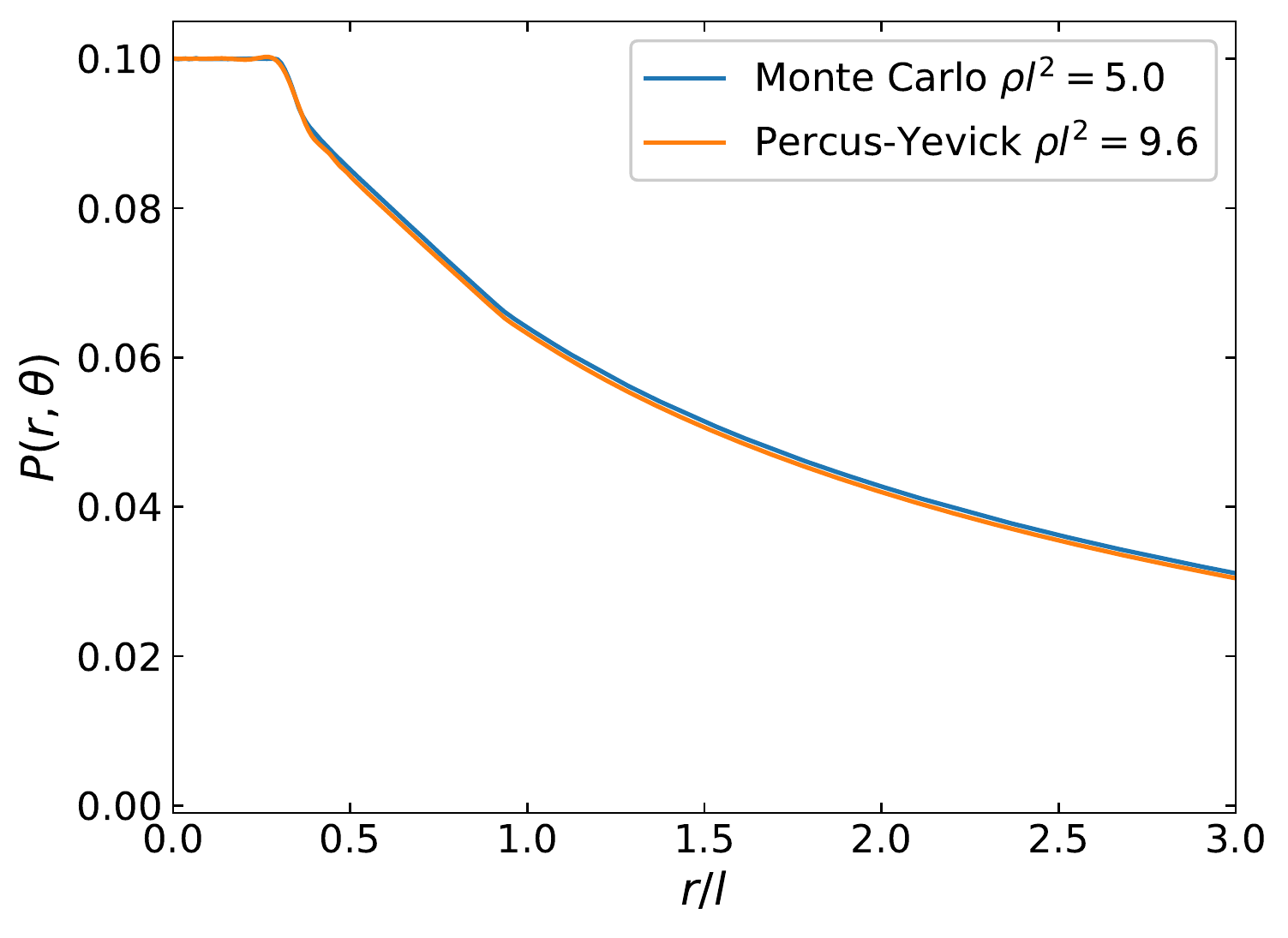}
	~
	\includegraphics[width=0.48\columnwidth]{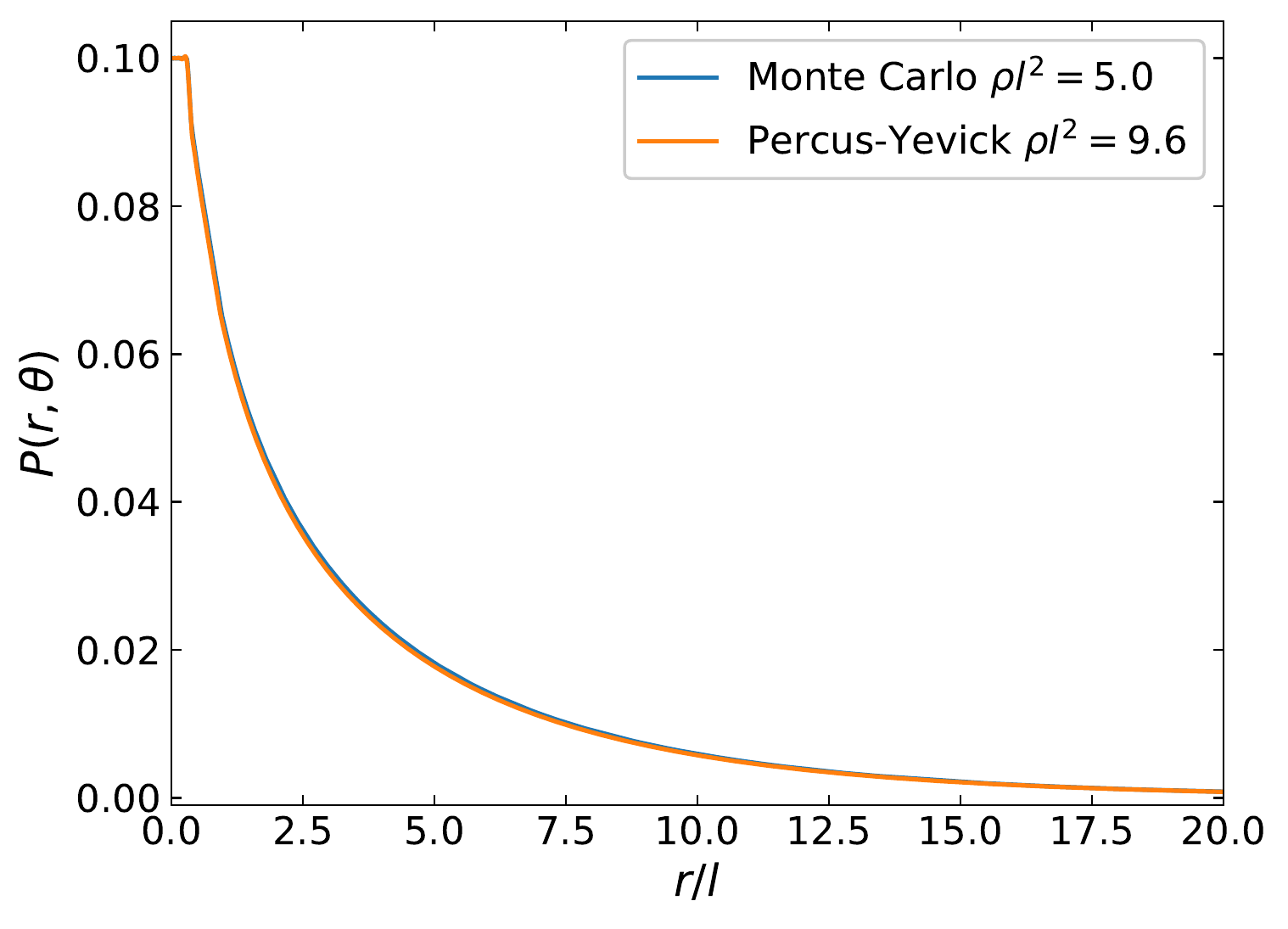}
	\caption{The normalized pair connectedness function $P(r,\vartheta)$ as function of the scaled center-to-center distance $r/l$. The pair connectedness function is averaged over the angle $\vartheta$ between the particles $36^\circ<\vartheta < 45^\circ$. Predictions due to cPY theory match Monte Carlo simulation results if evaluated at different densities, both at small and large separation.}
	\label{fig:pcftheta}
\end{figure}

While the state of affairs is unsatisfying for determining the percolation threshold, we can still obtain valuable information on connection probabilities below the percolation threshold. We show the average pair connectedness functions for two densities for both cPY approximation and from Monte Carlo simulations in Fig.~\ref{fig:pcf},
which show good agreement at low density, but less so at high density. Interestingly, evaluating cPY theory and Monte Carlo simulations at \textsl{different} densities does yield accurate results for the pair connectedness functions \footnote{This does not hold for the direct connectedness function, as it is still limited by the cPY approximation}. To highlight that this also holds for the non-averaged pair connectedness function, we plot the short and long range behaviour of the (normalized) pair connectedness function in Fig.~\ref{fig:pcftheta}, which for comparison with our Monte Carlo simulations is averaged over the angle-bin $36^\circ<\vartheta < 45^\circ$, where $\vartheta$ is now the angle between the major axes of the two particles. Here, we evaluate the cPY approximation at $\rho l^2 = 9.6$ with $n_\mathrm{max} = 20$ to properly capture the orientational information, whereas the Monte Carlo simulation results have been obtained for $\rho l^2 = 5$. Fig.~\ref{fig:pcftheta} is representative for all angles.

\begin{figure}[tb]
	\centering
	\includegraphics[width=0.9\columnwidth]{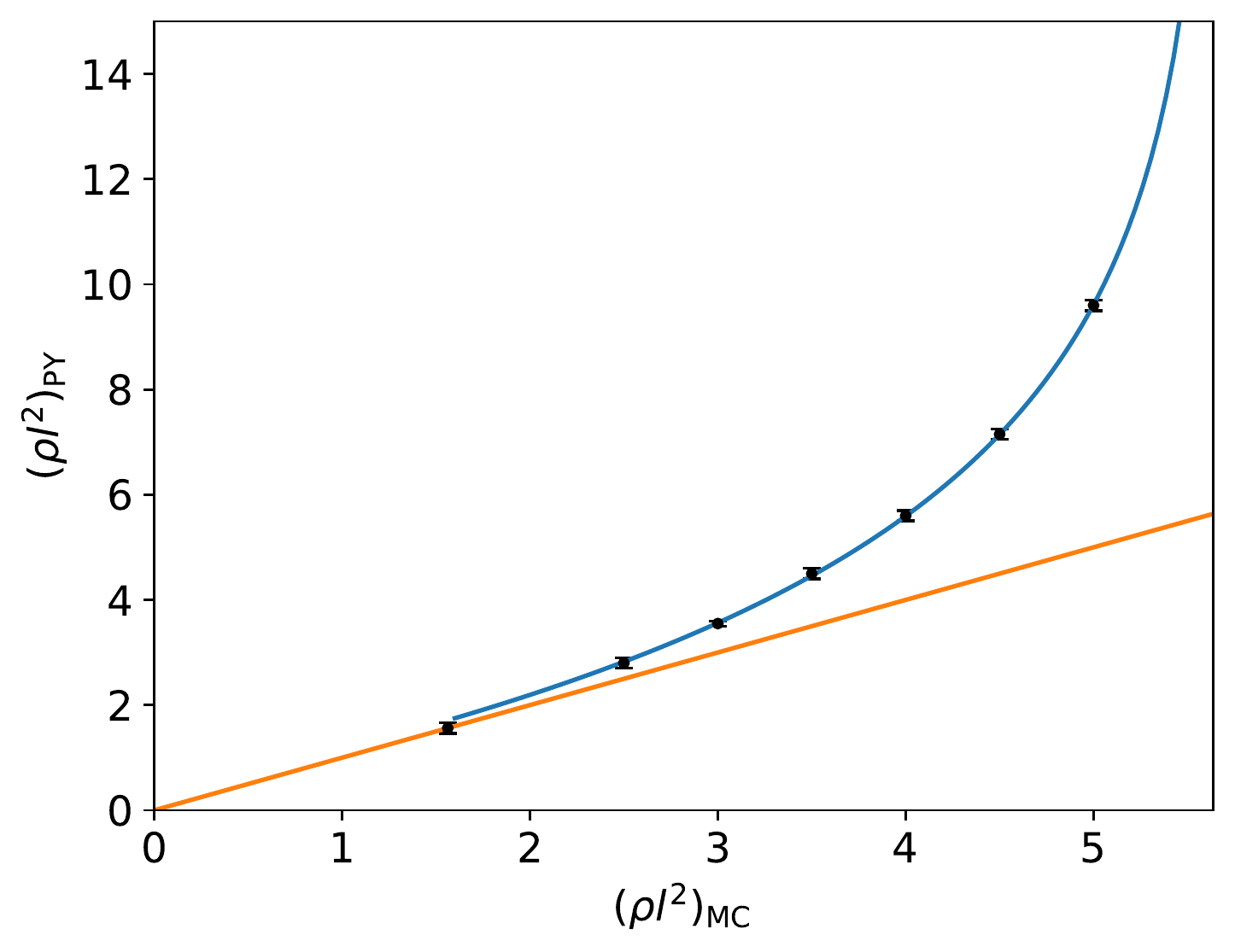}
	\caption{The density mapping of the pair connectedness function obtained from Percus-Yevick theory $P_\mathrm{PY}(r)$ on the pair connectedness function obtained from our Monte Carlo simulations $P_\mathrm{MC}(r)$, for ideal line segments of length $l$ in a two-dimensional model. The scaled density ($\rho l^2)_\mathrm{PY}$ for Percus-Yevick theory is plotted as function of the equivalent scaled density of Monte Carlo simulations $(\rho l^2)_\mathrm{MC}$. The orange line indicates $(\rho l^2)_\mathrm{PY} = (\rho l^2)_\mathrm{MC}$, the black dots indicate the density mapping with an uncertainty of $\Delta(\rho l^2) = 0.1$. The blue curve is the proposed fit Eq.~\eqref{eq:fit} with parameters $A = 7.69 \pm 0.01$ and $B = 4.26 \pm 0.02$.}
	\label{fig:denstmap}
\end{figure}

For the three-dimensional slender rod model, density renormalization schemes have been successfully used to extend the region of quantitative predictions of the second virial approximation from infinitely slender particles, to particles with an aspect ratio in the range of $L/D \sim \mathcal{O}(10^1)$ \cite{Schilling_2015}. Using the observation that the pair connectedness functions are in excellent agreement, as long as we evaluate our theoretical model at a different density, we propose an \textsl{ad-hoc} density renormalization approach as follows: (1) We match both the short-ranged and long-ranged behaviour of the pair connectedness function, (2) we assume that the deviation from the Monte Carlo results is due to it entering the critical region, \textit{i.e.}, the mean cluster size conforms to the scaling relation $S \sim |\rho - \rho_\mathrm{c}|^{-\gamma}$, and (3) we assume that the mean cluster size in cPY increases exponentially, which, from Fig.~\ref{fig:mcsrotinv}, seems to be a reasonable approximation. Under these considerations, we propose the following scaling relation
\begin{equation}\label{eq:fit}
\rho_\mathrm{PY} = A - B \ln\left(|\rho_\mathrm{MC} - \rho_\mathrm{c}|l^2\right),
\end{equation}
where $A$ and $B$ are fitting parameters and $\rho_\mathrm{c}$ is the percolation threshold. Since the  critical exponent $\gamma = 43/18$ associated with the mean cluster size is known, the parameter $B$ could in principle be calculated. Treating $\rho_\mathrm{c}$ as known, we obtain $A = 7.69 \pm 0.01$, and $B = 4.26 \pm 0.02$, which, as shown in Fig.~\ref{fig:denstmap}, is in excellent agreement. Since the same (near) exponential growth is also observed for ideal disks, we believe that such a method should be applicable for all ideal two-dimensional particle geometries \cite{deBruijn_2020}. 

While yielding excellent agreement with simulations, this density renormalization has two main drawbacks. Firstly, it relies on \textit{ad hoc} assumptions and requires direct input from simulations. Secondly, for non-ideal particles, \textit{e.g.}, obeying the cherry-pit model, this method cannot be straightforwardly applied. Practical application is therefore limited to simulation-assisted approaches.

\section*{NNCPT Mean Cluster Size}

In the main text we derived 
\begin{align}
	 P(r_{12}) &= f^+(r_{12}) + (1- f^+(r_{12})) \\ \label{eq:numericaleq} 
	&\times\int_{-\frac{1}{2}}^\frac{1}{2} \dd t' \int \dd \bs{r}_{3} \; \omega^+(\bs{r}_{3},l t')  l_{\mathrm{eff}}(l |t'|)  P(r_{32}) \; .
\end{align}
The integral over $\bs{r}_3$ is to be evaluated on the entire $\mathbb{R}^2$ plane and can thus be interchanged with the $t'$-integral. Defining 
\begin{align}
	C^+( \bs{r}_{3}) := 	\int_{-\frac{1}{2}}^\frac{1}{2} \dd t' \; \omega^+(\bs{r}_{3},l t')  l_{\mathrm{eff}}(l |t'|) \; ,
\end{align}
and recalling that we fixed line segment 1 in the origin we obtain
\begin{align}
	P(|\bs{r}_{2}|) &= f^+(|\bs{r}_{2}|) + (1- f^+(|\bs{r}_{2}|)) \\  \label{eq:splitter} 
 &\times \int \dd \bs{r}_{3} \; C^+(\bs{r}_{3})  P(|\bs{r}_2 - \bs{r}_3 |) \; .
\end{align}
The integral in the last equation is a two-dimensional convolution. The only important structural difference to the cOZ equation is the $|\bs{r}_{2}|$-dependent factor in front of the convolution. But, the support of $f^+(|\bs{r}_{2}|)$ is limited to a disk of radius $l$, so that for $|\bs{r}_{2}| > l$ the function in front of the convolution becomes unity. We split $P$ into two separate contributions:
\begin{align}
	P(|\bs{r}_2|) = \underbrace{P(|\bs{r}_2|)\Theta(|\bs{r}_2|-l)}_{P_{<l}(|\bs{r}_2|)} +  \underbrace{P(|\bs{r}_2|)\Theta(l-|\bs{r}_2|)}_{P_{>l}(|\bs{r}_2|)} \; ,
\end{align}
which applied to eq.~(\ref{eq:splitter}) yields two coupled integral equations:
\begin{eqnarray}
	P_{<l}(|\bs{r}_2|) &=&  f^+(|\bs{r}_{2}|) + (1- f^+(|\bs{r}_{2}|))  \\
	&&\int \dd \bs{r}_{3} \; C^+(\bs{r}_{3})   \nonumber
	\left[P_{<l}(|\bs{r}_2 - \bs{r}_3 |) +
	P_{>l}(|\bs{r}_2 - \bs{r}_3 |)\right] \\
		P_{>l}(|\bs{r}_2|) &=&  \nonumber
	\int \dd \bs{r}_{3} \; C^+(\bs{r}_{3})    
	\left[P_{<l}(|\bs{r}_2 - \bs{r}_3 |) +
	P_{>l}(|\bs{r}_2 - \bs{r}_3 |)\right] \label{eq:Plarger}
\end{eqnarray}
As the support of $P_{<l}$ is limited and $P(\bs{r}_2|) \leq 1$ necessarily, the percolation threshold depends exclusively on the properties of $P_{>l}$.  The impact of $P_{<l}$ in eq. (\ref{eq:Plarger}) can be accounted for by a new inhomogeneity $g$.
\begin{align}
		P_{>l}(|\bs{r}_2|) = \underbrace{\int \dd \bs{r}_{3} \; C^+(\bs{r}_{3})    P_{<l}(|\bs{r}_2 - \bs{r}_3 |) }_{g(|\bs{r}_2|)} \nonumber \\ +  \int \dd \bs{r}_{3} \; C^+(\bs{r}_{3})    
	P_{>l}(|\bs{r}_2 - \bs{r}_3 |) \label{eq:Plarger2} 
\end{align}
The mean cluster size consists of the contributions of both $P_{>l}$ and $P_{<l}$, although the latter is necessarily bounded by $\pi l^2$, because $P$ is bounded by 1. Thus, we can in good approximation neglect this contribution. The remaining part solves Eq.~(\ref{eq:Plarger2}) which is of the standard convolution type and thus can be solved in Fourier space
\begin{align}
\hat{P}(\bs{q}) =  \frac{\hat{g}(\bs{q})}{1-\hat{C}(\bs{q})}
\end{align}  

$\hat{g}(\bs{0})$ describes the integral of $g$ over the real plane. However, $g$ is a convolution itself and thus measure preserving. So $\hat{g}(\bs{0}) = \hat{C}^+(\bs{0}) \hat{P}_{<l}(\bs{0})$. But since $f^+(\bs{r}) \leq P_{<l}(\bs{r}) \leq 1$ for any $\bs{r}$, we know
\begin{align}
    \hat{f}^+(\bs{0}) \leq \hat{P}_{<l}(\bs{0}) \leq \pi l^2 \;.
\end{align}
Since $f^+$ does not depend on the density, $f^+(\bs{0})$ does only depend on $l$. Therefore, $\hat{P}_{<l}(\bs{0})$ is strongly constrained by density independent constants. As a consequence, for the sake of determining the mean cluster size the approximation $\hat{g}(\bs{q}) \approx \hat{C}^+(\bs{q})$ is well controlled. With this approximation eq.~(\ref{eq:Plarger2}) becomes an Ornstein-Zernike type equation, so that the mean cluster size can be computed via
\begin{align}
    S = \lim_{\bs{q} \rightarrow \bs{0}} \frac{1}{1-C^+(\bs{q})} \; .
\end{align}
This is the approximation of the mean cluster size from NNCPT which is depicted in Fig.~1 of the main text.
\begin{figure}[h!]
    \centering
    \includegraphics[width=0.8 \columnwidth]{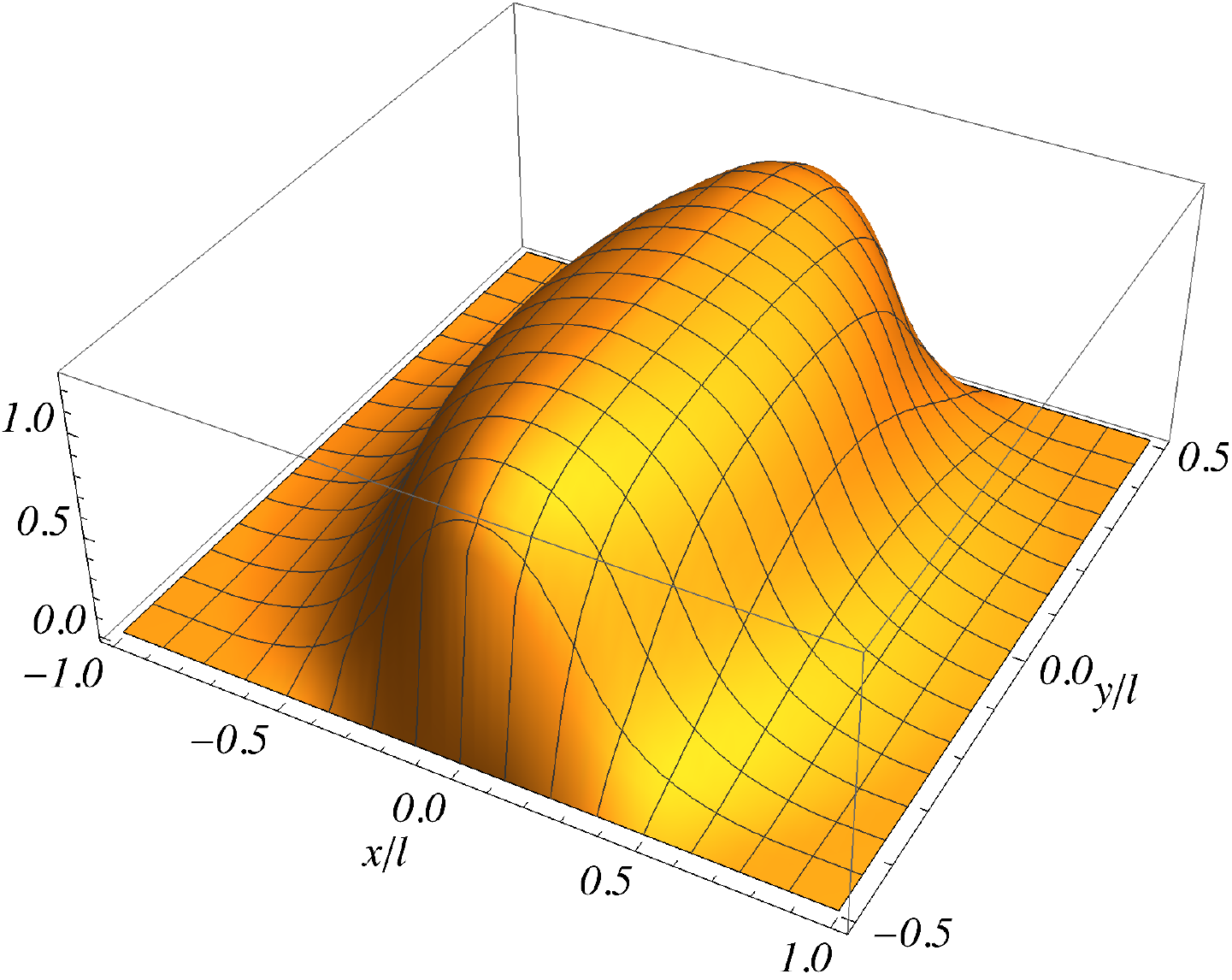}
    \caption{NNCPT closure: $C^+(\bs{r_3})$ for $\rho l^2=5$}
    \label{fig:nncpt_kernel}
\end{figure}

\end{document}